\documentclass[10pt, conference, letterpaper]{IEEEtran}
\usepackage{graphicx,subfig}
\usepackage{cite}
\usepackage{color}
\usepackage{bigstrut}
\usepackage{multirow}
\usepackage{amsmath}
\usepackage{url}
\usepackage{array}
\newcolumntype{L}[1]{>{\raggedright\let\newline\\\arraybackslash\hspace{0pt}}m{#1}}
\newcolumntype{C}[1]{>{\centering\let\newline\\\arraybackslash\hspace{0pt}}m{#1}}
\newcolumntype{R}[1]{>{\raggedleft\let\newline\\\arraybackslash\hspace{0pt}}m{#1}}
\usepackage{amssymb}
\DeclareMathOperator*{\argmax}{argmax}

\newcommand{\etal}{\textit{et al. }}

\begin{document}

\title{Radio Frequency Fingerprint Identification for Security in Low-Cost IoT Devices}

\author{
\IEEEauthorblockN{
Guanxiong~Shen\IEEEauthorrefmark{1},
Junqing~Zhang\IEEEauthorrefmark{1}\IEEEauthorrefmark{4},
Alan~Marshall\IEEEauthorrefmark{1},
Mikko~Valkama\IEEEauthorrefmark{2},
Joseph Cavallaro\IEEEauthorrefmark{3}
}

\IEEEauthorblockA{
\IEEEauthorrefmark{1}
Department of Electrical Engineering and Electronics, University of Liverpool, Liverpool, L69 3GJ, United Kingdom\\ 
Email: \{Guanxiong.Shen, junqing.zhang, alan.marshall\}@liverpool.ac.uk
}
\IEEEauthorblockA{
\IEEEauthorrefmark{2}
Department of Electrical Engineering,
Tampere University, 33720 Tampere, Finland
\\Email: mikko.valkama@tuni.fi
}
\IEEEauthorblockA{
\IEEEauthorrefmark{3}
Department of Electrical and Computer Engineering,
Rice University, Houston, USA\\ 
Email: cavallar@rice.edu
}
\IEEEauthorblockA{
\IEEEauthorrefmark{4}
Corresponding Author}
}
%\author{
%	\IEEEauthorblockN{
%		Author 1,
%		Author 2,
%	    Author 3,
%		Author 4,
%       Author 5,
%	and	Author 6
%	}
%
%}

\maketitle

\begin{abstract}

% \blue{Radio frequency fingerprint identification (RFFI) is a promising technique that can uniquely identify wireless devices.
% It is achieved by analyzing the received signal distortions caused by the intrinsic hardware impairments. RFFI is suitable to identify low-cost Internet of Things (IoT) devices as they possess abundant hardware impairments, and it does not increase the power consumption of the IoT end nodes at all. With the state-of-the-art deep learning technology, it can successfully classify IoT devices with high accuracy. However, there are still some obstacles in deep learning-based RFFI. First, most studies use convolutional neural network (CNN) for classification, but it cannot handle signals of variable length. Second, the transmission power of IoT devices is limited, resulting in low signal-to-noise ratio (SNR) of the received signal. However, the low SNR problem is rarely investigated in RFFI. In this paper, a transformer is used as the classifier, which can process signals of variable length. Augmentation is shown to be effective in improving low SNR RFFI performance. The performance of online, offline and no augmentation strategies are compared and the online one outperforms the others. A multi-packet inference protocol is further proposed to improve the classification accuracy in low SNR scenarios. We take LoRa as a case study and evaluate the system by classifying 10 commercial-off-the-shelf LoRa devices in various SNR conditions.}

Radio frequency fingerprint identification (RFFI) can uniquely classify wireless devices by analyzing the received signal distortions caused by the intrinsic hardware impairments. The state-of-the-art deep learning techniques such as convolutional neural network (CNN) have been adopted to classify IoT devices with high accuracy. However, deep learning-based RFFI requires input data of a fixed size. In addition, many IoT devices work in low signal-to-noise ratio (SNR) scenarios but the low SNR RFFI is rarely investigated. In this paper, the state-of-the-art transformer model is used as the classifier, which can process signals of variable length. Data augmentation is adopted to improve low SNR RFFI performance. A multi-packet inference approach is further proposed to improve the classification accuracy in low SNR scenarios. We take LoRa as a case study and evaluate the system by classifying 10 commercial-off-the-shelf LoRa devices in various SNR conditions. The online augmentation can boost the low SNR RFFI performance by up to 50\% and multi-packet inference can further increase it by over 20\%.

\end{abstract}
	

\section{Introduction}
%\IEEEPARstart{W}{ith} the evolution of Internet of Things (IoT) technology, its security has also attracted much more attention. Device authentication is a critical security mechanism that ensures the malicious devices cannot access to the IoT network. Current device authentication techniques usually rely on software addresses and cryptographic algorithms. However, some of them are so complex that cannot be implemented on low-cost IoT devices. A lightweight IoT authentication mechanism is desired.   

\IEEEPARstart{R}{adio} frequency fingerprint identification (RFFI) is a promising technique for authenticating Internet of things (IoT) devices. 
The analog front-end of wireless devices consists of hardware components such as oscillator, mixer, modulator, etc~\cite{valkama2010digital}. These components are rich in hardware impairments, whose specification parameters deviate from their nominal values.  
The impairments are unique among devices hence can be extracted as an identifier, working in a similar manner as human fingerprints~\cite{zhang2021radio}.
% Traditional RFFI techniques extract human-designed features, such as carrier frequency offset (CFO), statistical features, power amplifier non-linearity, differential constellation trace figure (DCTF), Hilbert-Huang spectrum, etc~\red{[refs]}.

As RFFI can be considered as a classification problem, deep learning techniques have been extensively exploited recently~\cite{yu2019robust,merchant2018deep,al2021deeplora,al2020exposing,soltani2020more,ozturk2020rf,roy2019rfal,qian2021specific,shen2021jsac,das2018deep}.
It uses a neural network to directly classify the received signals and predicts the device label without the need for manual feature extraction or expert knowledge.
Convolutional neural network (CNN) is the most widely adopted model thanks to its excellent feature extraction capability~\cite{yu2019robust,merchant2018deep,al2021deeplora,al2020exposing,soltani2020more,ozturk2020rf,roy2019rfal,qian2021specific,shen2021jsac}. However, CNN requires input signals of a fixed size~\cite{al2020exposing}. Therefore, some work uses fixed-length signals as model inputs~\cite{ozturk2020rf,roy2019rfal,qian2021specific,shen2021jsac,hanna2020open}, e.g., the preamble. This however limits the suitable data and rules out the payload part which is usually of variable lengths.
%Fixing the model input length is applicable for some protocols such as WiFi, since the length of WiFi preamble is fixed. However, not all protocols have a fixed-length preamble. For instance, LoRa preamble length can vary due to the special LoRaWAN adaptive data rate (ADR) mechanism.
Others proposed the slicing/splitting technique that divides the signal into equal-length segments~\cite{yu2019robust,merchant2018deep,al2020exposing,soltani2020more}. However, it is not clear whether some hardware features are lost, e.g., there might be temporal correlation hidden in the signal. A deep learning model that can process inputs of variable length is desired.

RFFI performance for IoT is also constrained by the low signal-to-noise ratio (SNR).
The transmission power of IoT end nodes should always be kept as minimum as possible to reduce power consumption.
One solution for low SNR RFFI is to design noise-resilient signal representations as model inputs.
Ozturk~\etal found that time-frequency data (spectrograms) is more resilient to noise than time-series data~\cite{ozturk2020rf}. Xing~\etal proposed a stacking algorithm for direct sequence spread spectrum (DSSS) systems to improve signal quality\cite{xing2018radio}.  
Alternatively, we can also obtain noise robustness by enhancing the capability of the RFFI model.
Data augmentation is a popular approach to train a channel-agnostic RFFI model~\cite{al2021deeplora,soltani2020more,merchant2019enhanced}, which can also be leveraged to improve the model noise robustness. However, specific analysis on the effect of data augmentation in low SNR scenarios is still missing.

In this paper, the state-of-the-art transformer model is used as the classifier, which can handle sequences of variable length~\cite{vaswani2017attention}.
% we employ the transformer model to classify LoRa devices with variable-length preambles and enhance classification performance under low SNR scenarios.
% Transformer can handle sequences of variable length~\cite{vaswani2017attention}
We leverage the data augmentation to train a noise-robust transformer, and compare the performance of different augmentation strategies. Finally, we propose a multi-packet inference approach, which can significantly improve the classification accuracy in low SNR scenarios.
% We take LoRa as a case study because the LoRaWAN adaptive data rate (ADR) mechanism changes the preamble length, making the model input size variable. 
We take LoRa as a case study since it faces both variable-length input and low SNR problems. However, the design methodology is applicable to any RFFI system with any wireless technology. Experimental evaluation is carried out using 10 commercial-off-the-shelf LoRa devices and a USRP N210 software-defined radio (SDR) platform. 
%The results show that online augmentation and multi-packet inference protocol can effectively improve the performance of low-SNR RFFI.
% The transformer leverages the mechanism of attention, which was originally designed for sequential data input~\cite{vaswani2017attention}. It has been extensively applied in speech recognition~\cite{dong2018speech}, image classification~\cite{dosovitskiy2020image}, natural language processing~\cite{devlin2018bert} tasks and demonstrated strong learning ability.
Our contributions are listed as follows:
\begin{itemize}
    \item We design a transformer model to classify LoRa devices with variable-length preambles. %The proposed transformer model is able to process sequential data of variable length, which is more suitable in LoRaWAN applications. 
    To the best knowledge of the authors, it is the first time to apply the state-of-the-art transformer model into RFFI.
    \item We investigate the effect of data augmentation on enhancing RFFI performance, especially in low SNR scenarios. The transformer models are trained with online, offline and no augmentation strategies and online augmentation performs the best. Specifically, online augmentation improves the classification accuracy by 40\% at 15~dB compared to the model without augmentation.		
    \item We propose a multiple-packet inference method that can significantly improve the system performance in low SNR scenarios. It can boost the classification accuracy from 60\% to 90\% at 10~dB.
    
\end{itemize}

\section{Preliminaries}\label{sec:preliminary}
% The LoRaWAN adaptive data rate (ADR) mechanism must be considered in the design of LoRa-RFFI system. More specifically, it changes 

\subsection{LoRa Signal Modelling}
LoRa uses chirp spread spectrum (CSS) as the physical layer modulation technique. There are several basic up-chirps at the beginning of a LoRa packet, named preambles. A baseband LoRa preamble is given as
\begin{equation} 
	s(t) = A e^{j(-\pi Bt + \pi \frac{B}{T} t^2)} \quad (0 \leq t \leq \frac{2^{SF}}{B}), 
\end{equation}
where $A$, $B$ and $SF$ denote signal amplitude, bandwidth and spreading factor, respectively.
% $T$ is the duration of a LoRa symbol, given as
% \begin{equation}\label{equ:preamble_time}
%     T = \frac{2^{SF}}{B}, 
% \end{equation}
The bandwidth can be 125~kHz, 250 kHz and 500 kHz and the SF ranges from 7 to 12. 

\subsection{LoRaWAN Adaptive Data Rate}\label{sec:ADR}
The adaptive data rate (ADR) mechanism is adopted in LoRaWAN, which allows LoRa end-nodes to optimize the spreading factor, bandwidth and transmission power on the fly.
The LoRa symbol duration is affected by bandwidth and spreading factor. Therefore, LoRaWAN ADR mechanism makes the duration of packet preamble $s(t)$ variable.

\subsection{Channel Independent Spectrogram}
As LoRa uses CSS modulation, its frequency changes over time. Hence a spectrogram can be obtained via short-time Fourier transform (STFT) to reveal its frequency variation in the time-frequency domain. A channel independent spectrogram, $\mathbf{S}$, can be further generated to mitigate channel effects, with each matrix element mathematically given as 
\begin{align}
		S_{k,m} &= 10\log_{10} \Big(\left | \frac{\sum_{n=0}^{N-1} s[n] w[n-mR]e^{-j2\pi \frac{k}{N} n}}{\sum_{n=0}^{N-1} s[n] w[n-(m-1)R]e^{-j2\pi \frac{k}{N} n}}  \right |^2 \Big)\nonumber\\ 
		&\mbox{for}\ k = 1,2,..., N\ \mbox{and}\ m = 1,..., M-1,
		\label{equ:channel_ind_spectrogram}
\end{align}
where $s[n]$ is the digital sample of $s(t)$, $w[n]$ is a rectangular window of length $N$, and $R$ is the hop size. 
In this paper, $N$ and $R$ are always set to 64 and 32, respectively. Please refer to~\cite{shen2021towards} for the detailed derivation of the channel independent spectrogram.

%\blue{Another benefit of leveraging channel independent spectrograms as model inputs is that time-frequency data is more robust to noise than time-domain data~\cite{ozturk2020rf}.} 

\begin{figure}[!t]
	\centering
	\includegraphics[width=3.4in]{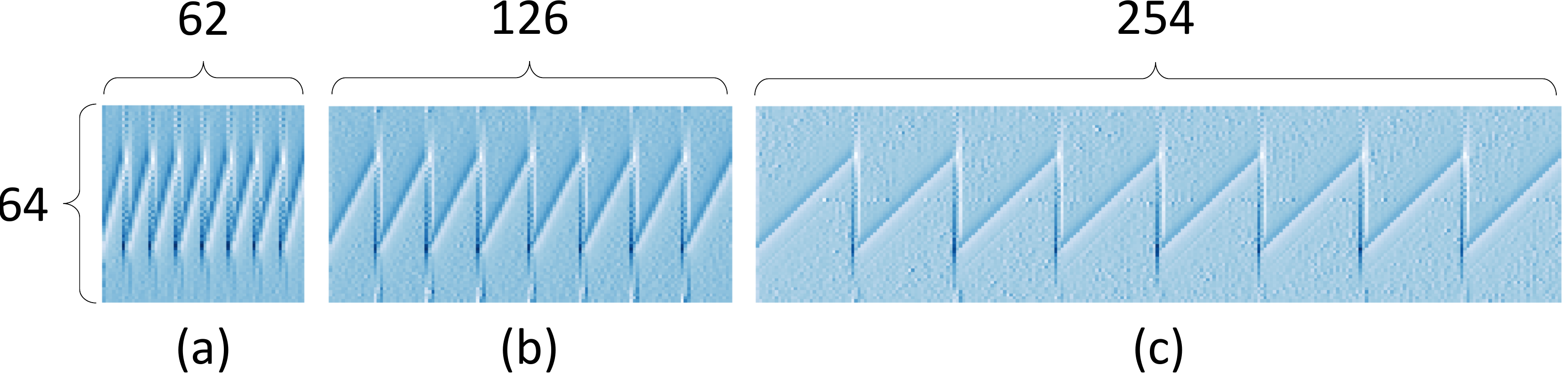}
	\caption{Channel independent spectrograms under different SF settings. (a) SF = 7. (b) SF = 8. (c) SF = 9.}
	\label{fig:channel_ind_spectrogram}
\end{figure}

LoRa ADR mechanism affects the size of $\mathbf{S}$. As shown in~(\ref{equ:channel_ind_spectrogram}), the height of $\mathbf{S}$, $N$, is equal to the length of window function $w[n]$. It can be configured the same for all SFs. However, the width of $\mathbf{S}$, $(M-1)$, is related to the length of $s[n]$, which is affected by the SF configuration. Therefore, LoRaWAN ADR mechanism makes the width of $\mathbf{S}$ variable.
% \blue{explain the height refers to the bandwidth so they are the same. the width is related to the time}
% LoRa-RFFI systems are thus required to process signals under different SF and BW configurations.
% Among them, the adjustment of SF will lead to the change of symbol/signal length. Therefore, LoRa-RFFI systems are required to process signals under different SF configurations.
% As mentioned in the last subsection, the LoRa preamble length is not fixed due to the ADR mechanism.
% As described in (\ref{equ:channel_ind_spectrogram}), a change in the length of $s[n]$ results in channel independent spectrograms $\mathbf{S}_{k,m}$ of different sizes.
Fig.~\ref{fig:channel_ind_spectrogram} shows channel independent spectrograms when SF is configured to 7, 8 and 9, respectively.
They have the same height but different widths. When SF is set to 9, the channel independent spectrogram is about four times as wide as when SF is set to 7. This demonstrates the input size of the deep learning model is not fixed in LoRa-RFFI. Thus a neural network that can process signals of variable lengths is desired.

\section{RFFI System}

\subsection{Overview}

\begin{figure}[!t]
	\centering
	\includegraphics[width=3.4in]{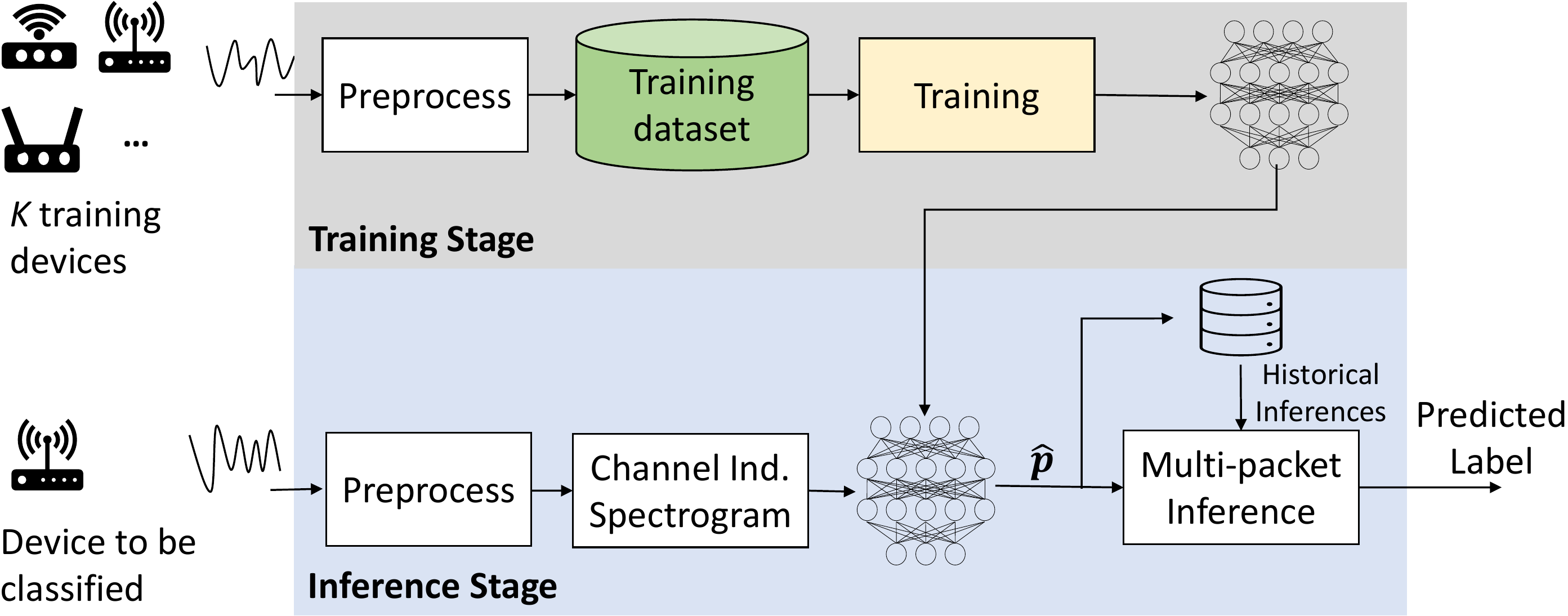}
	\caption{System overview.}
	\label{fig:system_overview}
\end{figure}

As shown in Fig.~\ref{fig:system_overview}, the proposed RFFI system involves two stages, namely training and inference. First, we collect extensive labelled LoRa packets from $K$ training devices, preprocess and store them as the training dataset. 
We specifically adopt transformer as the classifier, which is capable to process variable-length sequential data. Data augmentation is adopted during training.
Once the transformer training is completed, it can classify the newly received LoRa packet, named the inference stage. The LoRa packet is preprocessed and converted to the channel independent spectrogram to mitigate channel effects. Then it is fed into the trained transformer model and inference will be made.
In low SNR scenarios, we can leverage multiple packets for joint inference to obtain a more accurate prediction.

% One considerable advantage of transformer is that it can process input of variable length, which is more flexible than CNN. As introduced in Section~\ref{sec:preliminary}, the LoRaWAN ADR mechanism results in channel independent spectrograms of different widths. When CNN is used as the classifier, an independent CNN needs to be trained for each SF setting. In contrast, one single transformer model can handle all SF configurations.
% We employ the state-of-the-art transformer model to classify LoRa devices because of its ability to handle sequences of variable length and excellent feature extraction capabilities. The transformer leverages the mechanism of attention, which was originally designed for sequential data input~\cite{vaswani2017attention}. It has been extensively applied in speech recognition~\cite{dong2018speech}, image classification~\cite{dosovitskiy2020image}, natural language processing~\cite{devlin2018bert} tasks and demonstrated strong learning ability.
% \jz{move this to the system overview}

\subsection{Preprocessing}
As shown in Fig.~\ref{fig:system_overview}, the RFFI system needs to preprocess the received LoRa packet, both in the training and inference stages. We use the same preprocessing algorithms as in~\cite{shen2021towards}.

We employ the packet detection and synchronization algorithms to capture LoRa transmissions. After that, only the packet preamble part is kept for classification to prevent the model from learning protocol-specific information.
Then,  carrier frequency offset (CFO) compensation is carried out since CFO is sensitive to temperature variation and reduces the system stability.
Finally, the signal is normalized by dividing by its root mean square, to prevent the model from classifying devices relying on the received signal strength. 

%It involves the following four steps:

%\subsubsection{Synchronization and Preamble Extraction}
%We employ the packet detection and synchronization algorithms introduced in~\cite{robyns2018multi,shen2021infocom,shen2021jsac} to capture LoRa transmissions. After that, only the packet preamble part is reserved for classification to prevent the model from learning protocol-specific information.

%\subsubsection{Carrier Frequency Offset Compensation}
%The carrier frequency offset (CFO) refers to the frequency deviation from the standard one. It can contribute to the classification but must be compensated in deep learning-based RFFI systems. One reason is that CFO is sensitive to temperature variations, making the RFFI system not stable over time~\cite{andrews2019extensions,shen2021jsac,shen2021infocom}. The second is a deep learning model trained with uncompensated signals considers the CFO as the primary feature, making it fragile to CFO spoof attacks~\cite{merchant2019enhanced,yu2019robust}.
%We employ the LoRa CFO compensation algorithm proposed in~\cite{shen2021jsac} to avoid the above-mentioned problems.

%\subsubsection{Power Normalization}
%The signal is finally normalized by dividing its root mean square, to prevent the model from classifying devices relied on the received signal strength. 

\subsection{Training with Augmentation}\label{sec:training_pipeline}
In the training stage, we will collect a large number of LoRa packets, preprocess and store them as a training dataset. Note that we need to collect LoRa packets transmitted with various SF settings for training so that the transformer can handle different SFs in the inference stage.

\begin{figure}[!t]
	\centering
	\includegraphics[width=3in]{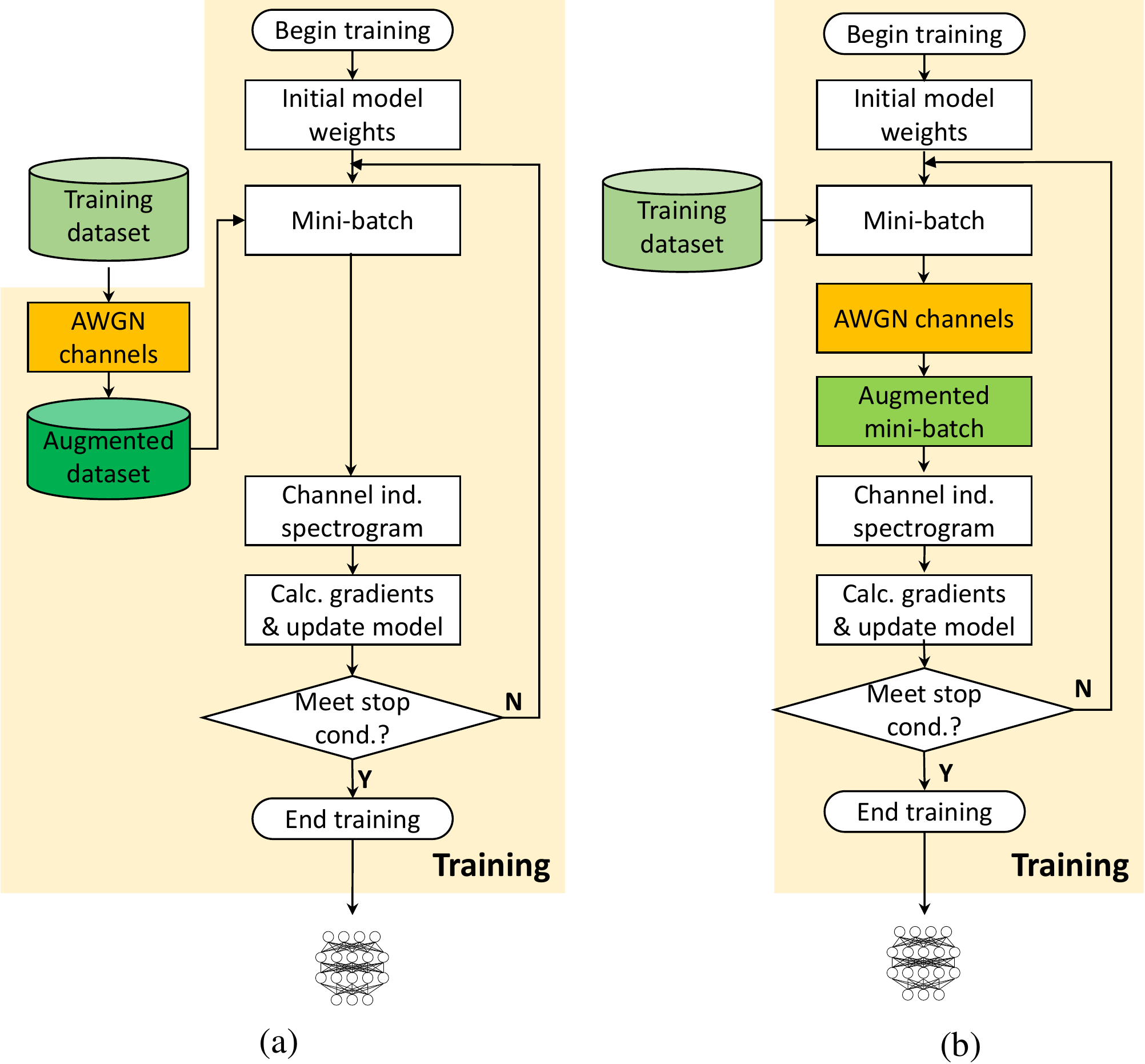}
	\caption{Training pipelines. (a) Offline augmentation. (b) Online augmentation.}
	\label{fig:training_pipelines}
\end{figure}

We investigated two training pipelines with different augmentation strategies, namely offline and online augmentation. The details are shown in Fig.~\ref{fig:training_pipelines}.
Data augmentation is usually employed in the training stage of an RFFI system to increase the model robustness against channel variations~\cite{shen2021towards,merchant2019enhanced,al2021deeplora,soltani2020more}. In this paper, we focus on different SNR scenarios therefore additive white Gaussian noise (AWGN) channel model is used for augmentation.

% \begin{figure}[!t]
% 	\centering
% 	\subfloat[]{\includegraphics[width=3.4in]{Pictures/offline_aug.pdf}
% 		\label{}}
		
% 	\subfloat[]{\includegraphics[width=1.9in]{Pictures/online_aug.pdf}
% 		\label{}}

% 	\caption{Training pipelines. (a) Offline augmentation. (b) Online augmentation.}
% 	\label{fig:augmentation}
% \end{figure}

\subsubsection{Offline Augmentation}
The offline augmentation is performed on the original training dataset. We pass the training samples through AWGN channels and store the outputs as the augmented training dataset. 
% Deep learning models are usually trained by mini-batch gradient descent optimization algorithms. 

At the beginning of the training, a number of samples are randomly selected from the augmented training dataset to form a mini-batch. The IQ samples in the mini-batch are then converted into channel independent spectrograms, which are input into the transformer. The gradients are calculated to update the model weights. This update process continues until training stop conditions are met.
% \blue{Note that each mini-batch contains signals with the same SF configuration, because all the samples in the mini-batch are required to be of the same length.}    
% Note that each mini-batch can only contain signals of exactly the same length, making each update step optimizes for a specific SF. Therefore, a switcher is used when selecting mini-batches to ensure that the training process covers all SFs.

\subsubsection{Online Augmentation}
The online augmentation is also known as augmentation on the fly. Different from the offline one, online augmentation is performed on the mini-batches. As shown in Fig.~\ref{fig:training_pipelines}, the mini-batch is selected from the training dataset and fed into AWGN channels, obtaining an augmented mini-batch. Other training steps remain the same as the offline augmentation. In online augmentation, the model is trained with (steps $\times$ mini-batch size) noisy signals.

\subsection{Transformer Architecture}
Transformer is originally designed as a sequence transduction model~\cite{vaswani2017attention}. Its main components, the multi-head attention layer and the point-wise feed forward layer can handle the variable-length sequences, and the input of dense layer is always a fixed-length vector. Therefore, the transformer can process sequences of any length during inference.

\begin{figure}[!t]
	\centering
	\includegraphics[width=2.5in]{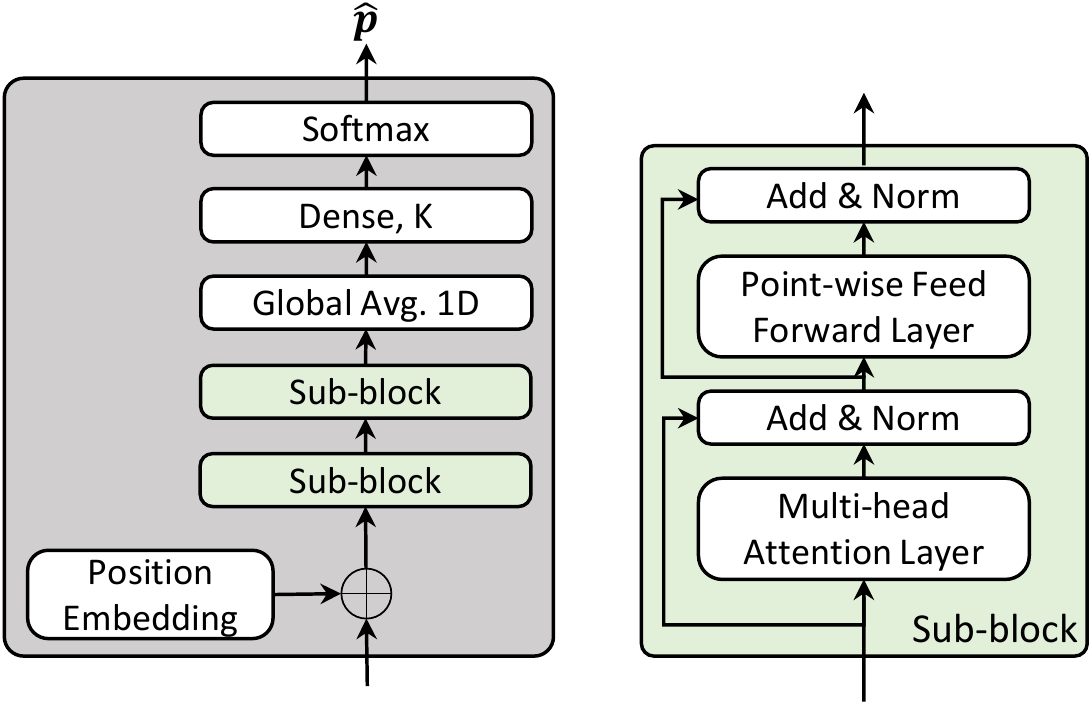}
	\caption{Transformer architecture.}
	\label{fig:transformer}
\end{figure}

The architecture of the used transformer is shown in Fig.~\ref{fig:transformer}. 
The model inputs are channel independent spectrograms introduced in Section~\ref{sec:preliminary}. Physically, the channel independent spectrogram shows time on the horizontal axis and frequency on the vertical axis. When it acts as input to the transformer, its columns (frames) are considered as time steps and frequency amplitudes are processed as features. First, position embedding is added to the model input to enable the model to learn temporal correlations. Then it is fed into two sub-blocks for feature extraction. Each sub-block contains a multi-head attention layer and a point-wise feed forward layer. The skip-connection and layer normalization are also leveraged. Then a global average pooling 1D layer is used to average the output of the second sub-block, which is finally connected to a $K$-neuron dense layer with softmax activation. The transformer is trained with the cross-entropy loss.

\subsection{Multi-packet Inference}
Once the training is completed, the transformer can act as a classifier to infer device identity. The received packet is preprocessed and converted to the channel independent spectrogram. Then it is fed into the well-trained transformer and a list of probabilities, $\mathbf{\hat{p}}$, will be returned by the softmax layer. $\mathbf{\hat{p}}$ can be regarded as confidence levels over $K$ devices. 

Multi-packet inference refers to making decisions with multiple LoRa transmissions. More specifically, we average the inference $\mathbf{\hat{p}}$ with $(N_{pkt}-1)$ historical inferences to derive a merged prediction $\mathbf{\hat{p}}{'}$, mathematically expressed as
\begin{equation}\label{equ:multi_pkt}
    \mathbf{\hat{p}}{'} = \frac{1}{N_{pkt}}\sum_{n=1}^{N_{pkt}}  \mathbf{\hat{p}}^{n}.
\end{equation}
where $\mathbf{\hat{p}}^{n}$ is the prediction from the $n^{th}$ packet.
Then the predicted label can be derived by selecting the index with the highest probability, which is formulated as 
\begin{equation}
    label = \mathop{\argmax}_{k}(\mathbf{\hat{p}{'}}).
\end{equation}

%According to the experimental results, the multi-packet inference is particularly effective in low SNR scenarios. However, in the high SNR scenarios, this protocol may not be used to save computing resources. A more detailed discussion can be found in Section~\ref{sec:experiment}.

% can significantly improve system performance in low SNR scenarios. While the performance benefits at relatively high SNRs is limited.

% \begin{equation}
%     \mathbf{\hat{p}}^{fused} = \sum_{j=1}^{J} \frac{\gamma^j}{\sum_{j=1}^{J}\gamma^j}  \mathbf{\hat{p}}^j,
% \end{equation}

\section{Experimental Evaluation}\label{sec:experiment}
\subsection{Experimental Settings}
\subsubsection{Data Collection Settings}
\begin{figure}[!t]
	\centering
	\subfloat[]{\includegraphics[width=1.7in]{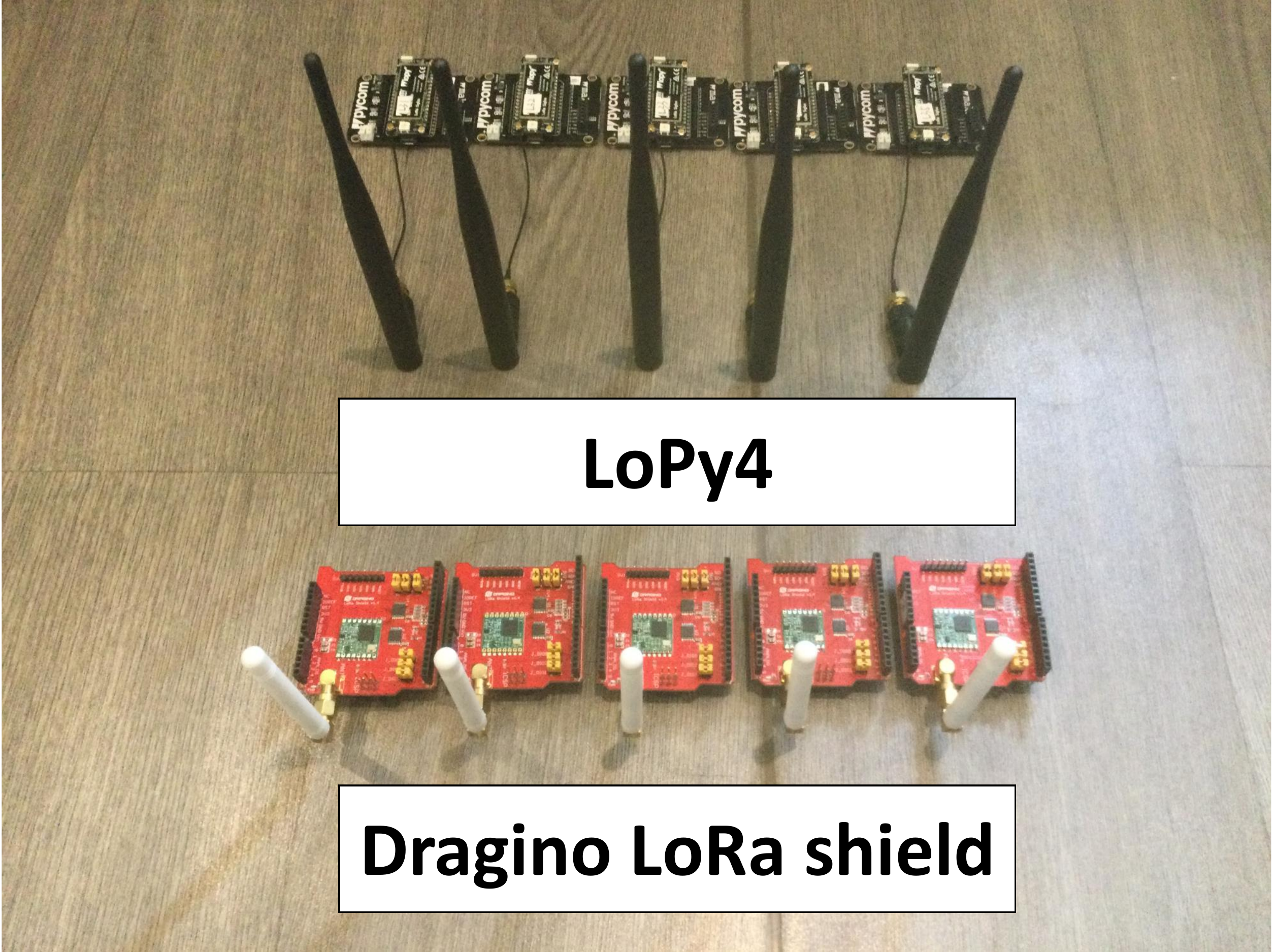}
		\label{}}
	\subfloat[]{\includegraphics[width=1.7in]{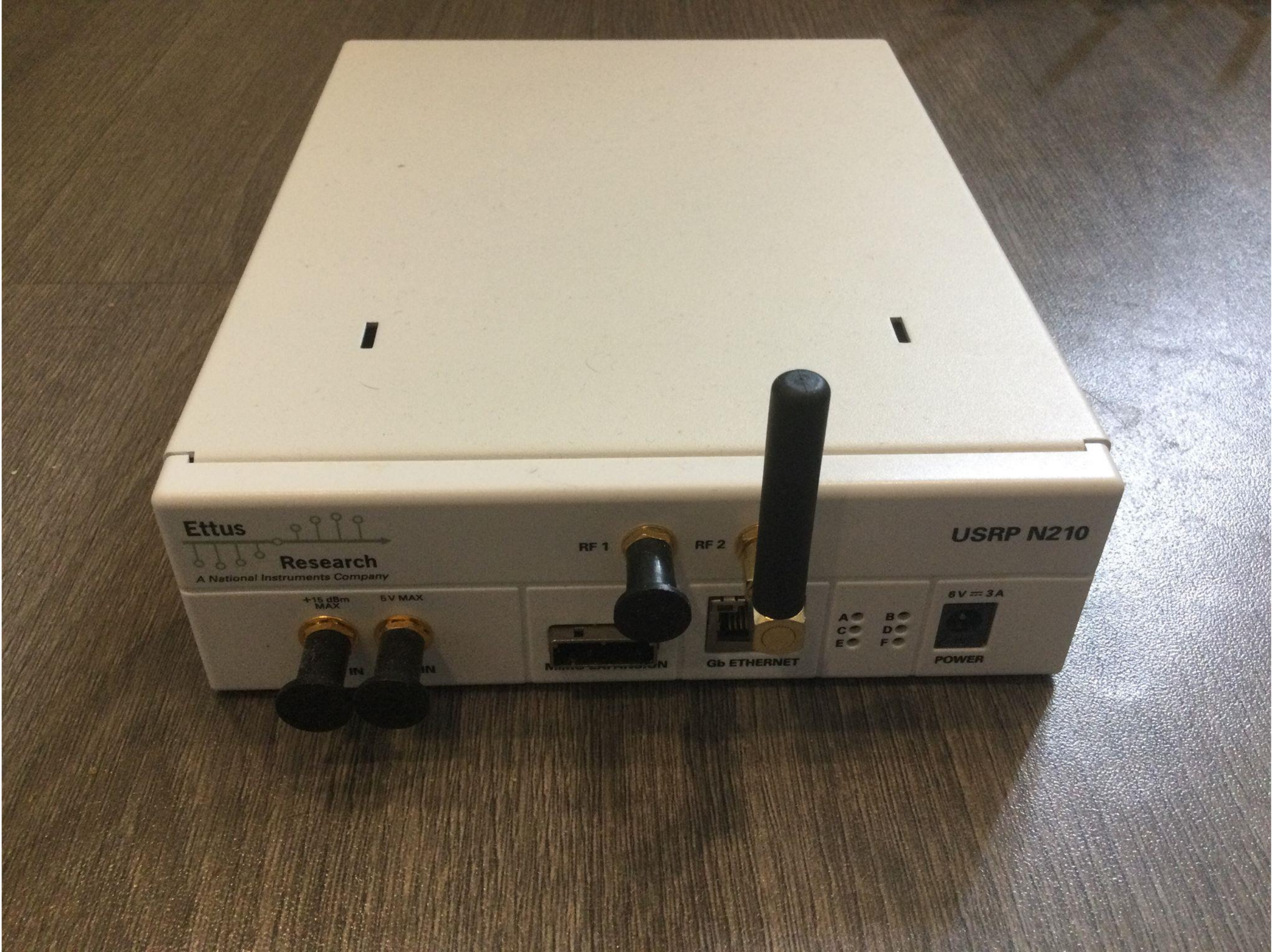}
		\label{}}
	\caption{Experimental devices. (a) LoRa DUTs. (b) USRP N210.}
	\label{fig:exp_dev}
\end{figure}

10 LoRa device under tests (DUTs) are employed. As shown in Fig.~\ref{fig:exp_dev}, five of them are LoPy4 and the others are Dragino LoRa shields, and they all use Semtech SX1276 chipsets.
All LoRa DUTs are configured with a bandwidth of 125~kHz, a carrier frequency of 868.1~MHz and a 0.5~second transmission interval. A USRP N210 platform is adopted as the receiver, and its sampling rate is set to 250~kHz. The LoRa DUTs and USRP N210 are placed half a meter apart with no obstacles between them (line-of-sight). Data collection was conducted device by device. Each LoRa DUT, in turn, set its SF to 7, 8, 9 and transmitted 3,000 packets per SF. 

\subsubsection{Dataset Description}
The experimental dataset contains 90,000 packets from 10 DUTs under three SF configurations. 75,000 of them (25,000 for each SF) are used for training, and the rest 15,000 (5,000 for each SF) for testing.  The SNR of all the received packets is estimated to be around 70~dB. Therefore, we treat them as noiseless signals and add AWGN to them to simulate different SNR conditions during the test.

\subsubsection{Transformer Training Details}
The transformer is trained according to the pipelines described in Section~\ref{sec:training_pipeline}. The SNR uniformly ranges from 0~dB to 40~dB  when generating the AWGN channels for both online and offline augmentation. 10\% training packets are randomly selected for validation. RMSprop is used as the optimizer and the mini-batch size is set to 32. A learning rate scheduler is adopted. The initial learning rate is 0.001 and is reduced by a factor of 0.2 every time the validation loss stops decreasing for 10 epochs. The training stops when validation loss plateaus for 20 epochs.
The transformer is implemented with Tensorflow library, and trained using a GPU of NVIDIA GeForce GTX 1660.

\subsection{Results}
\begin{figure}[!t]
	\centering
	\includegraphics[width=3.4in]{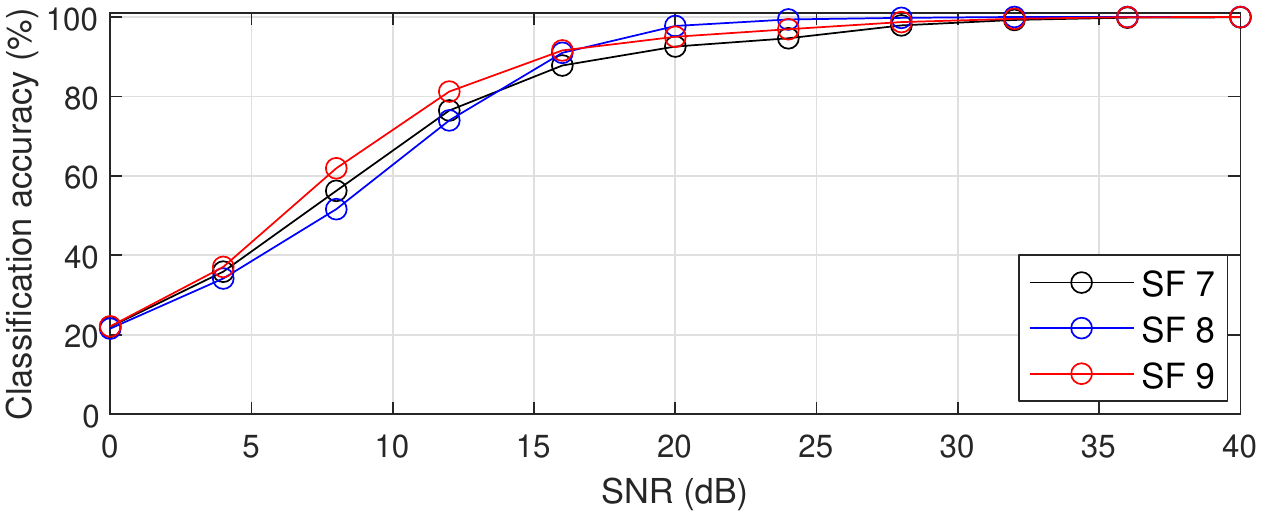}
	\caption{Classification accuracy under different SFs. The transformer is trained with SF 7, 8, 9 signals using online augmentation. Inference with SF 7, 8, 9 signals without the use of multi-packet method.}
	\label{fig:spreading_factor}
\end{figure}

\begin{figure}[!t]
	\centering
	\includegraphics[width=3.4in]{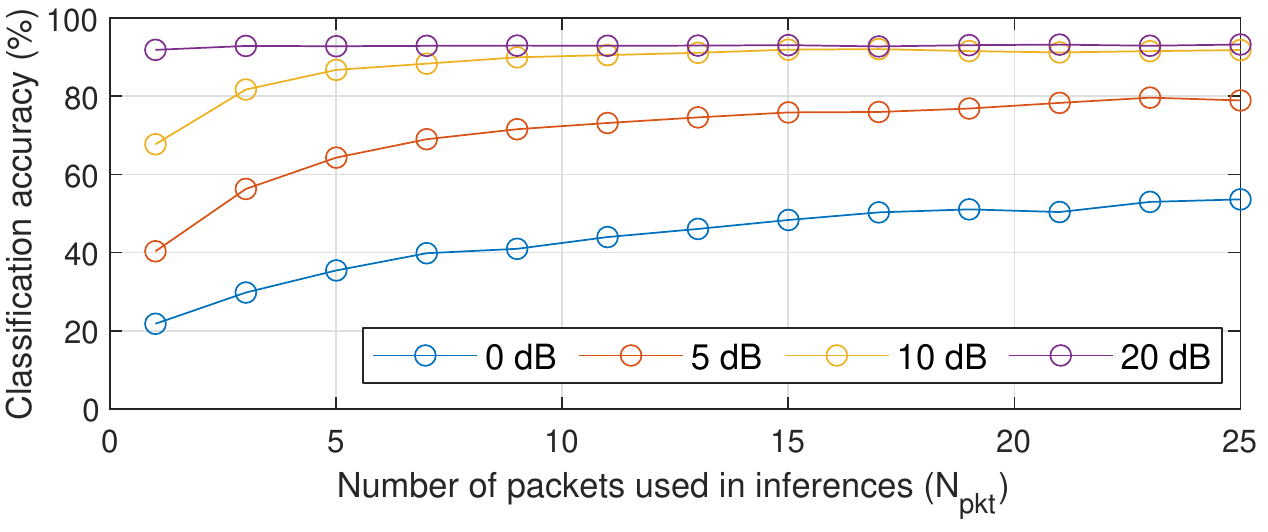}
	\caption{Effect of multi-packet inference method. The transformer is trained with SF 7, 8, 9 signals using online augmentation. Inference with SF 7 signals.}
	\label{fig:multi_pkt_inference}
\end{figure}

The transformer can process channel independent spectrograms under different SF configurations (Fig.~\ref{fig:channel_ind_spectrogram}). The classification results are shown in Fig.~\ref{fig:spreading_factor}. It demonstrates that the system performance is nearly perfect when SNR is over 30~dB. 
% At 20~dB, the accuracy can remain above 90\%. 
The accuracy is reduced to 60\% at 10 dB, but the multi-packet inference can increase it to 90\%, which will be shown in Fig.~\ref{fig:multi_pkt_inference}.
There is no evident performance gap among different SF settings. Therefore, the following experiments only use SF~7 signals during inference for clarity. 

\begin{figure}[!t]
	\centering
	\includegraphics[width=3.4in]{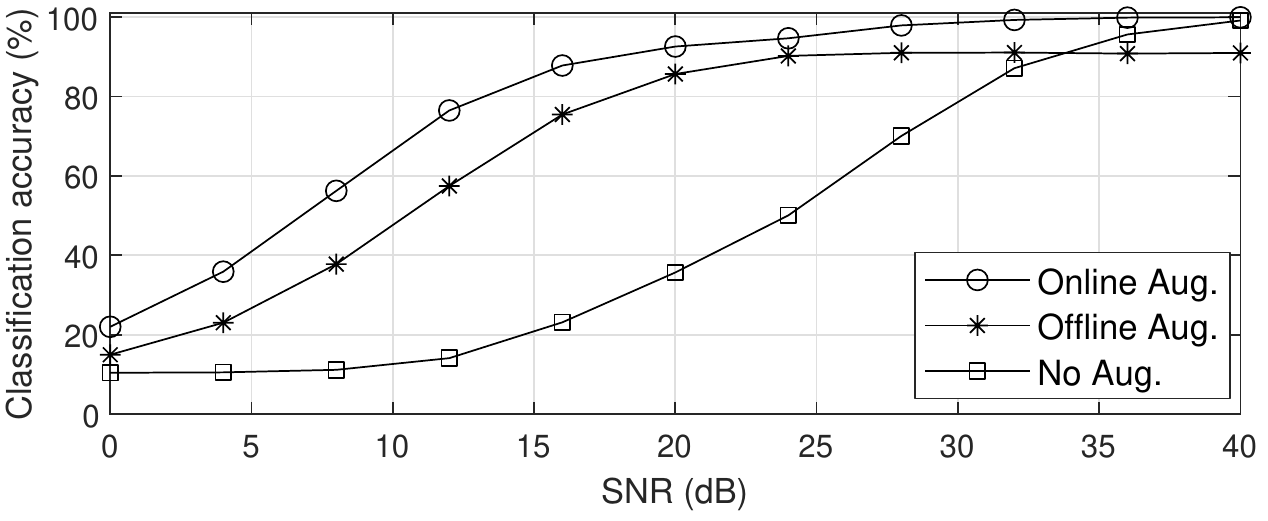}
	\caption{Comparison among augmentation strategies. The three transformers are trained with SF 7, 8, 9 signals using online, offline and no augmentation strategies, respectively. Inference with SF 7 signals without the use of multi-packet method.}
	\label{fig:augmentation_comparison}
\end{figure}

We trained three transformers using online, offline and no augmentation strategies, respectively. Their performance at different SNRs is shown in Fig.~\ref{fig:augmentation_comparison}. 
As illustrated in the figure, the transformer trained with online augmentation outperforms the one trained with offline augmentation. Because the transformer can learn more noisy signals during the online augmentation compared to the offline case. Specifically, the training of the online augmentation stops after 260,000 steps. Therefore, it has learned $260,000\times32$ (mini-batch size) noisy signals. In contrast, offline augmentation can only provide 75,000 (training set size) noisy signals to the transformer for learning.
It can also be observed that the transformer trained without augmentation performs the worst in low SNR scenarios. This demonstrates data augmentation must be employed during training to achieve noise robustness.

The proposed multi-packet inference approach can significantly improve RFFI performance, particularly in low SNR scenarios. The effect of packet number on classification accuracy is studied at four SNRs, and the results are given in Fig.~\ref{fig:multi_pkt_inference}. 
As shown in the figure, the multi-packet inference can increase the classification accuracy by over 20\% at 0~dB, 5~dB and 10~dB, while obtaining marginal improvement at 20~dB. This demonstrates multi-packet inference can be adopted in low SNR scenarios to effectively improve system performance.
Furthermore, we can find that the classification accuracy gradually improves as the number of packets increases. However, the improvement is relatively limited after the number of packets exceeds 10. A trade-off should be considered as involving more packets also leads to higher complexity and requires more space to store historical inferences.

\subsection{Complexity Analysis}
\subsubsection{Training Overhead} The online, offline, and no augmentation strategies require 200~mins, 80~mins and 140~mins to train the model, respectively. Online augmentation leads to the best performance, but also requires more training resources. Offline augmentation requires the least training time and can still maintain acceptable noise robustness. Note that training overhead is not critical for RFFI as the transformer can be trained on a server with strong computing capabilities and then distributed to the edge for inference.

\subsubsection{Inference Overhead} The inference time is 28.7~ms, 30.3~ms and 33.8~ms for SF 7, 8, 9, respectively. Note that this is limited by the deep learning library and can be accelerated. The trained transformer model contains 348,938 parameters in total, which is affordable for most embedded platforms.

The multi-packet inference protocol requires few storage and computing resources. We  need to store $(N_{pkt}-1)$ historical inferences $\mathbf{\hat{p}}$ for each LoRa device. Therefore, the historical inference database contains $K\times(N_{pkt}-1)$ vectors and each vector has $K$ probabilities. Therefore, there are $K^2\times(N_{pkt}-1)$ float numbers in total. As described in (\ref{equ:multi_pkt}), the multi-packet inference only requires an additional averaging operation, which is low-complexity.

% \subsubsection{Model Size} 

% \subsubsection{Cost of Multi-packet Inference}

\section{Conclusion}
This paper proposes a transformer-based RFFI system to tackle input signals with variable sizes.  LoRaWAN ADR mechanism results in LoRa preambles with variable lengths, requiring the model to be able to process inputs of different sizes. The transformer is capable of such tasks and adopted in this paper. The accuracy of LoRa-RFFI suffers in low SNR cases. Data augmentation and multi-packet inference are leveraged to enhance classification accuracy in low SNR scenarios.
We trained the transformer models with online, offline and no augmentation strategies and found that online augmentation is the most effective in enhancing model noise robustness. Furthermore, we designed a multi-packet inference approach that can considerably improve the system performance in low SNR scenarios. Experimental evaluation involving 10 LoRa devices and a USRP N210 SDR were carried out. %The results show that online augmentation and multi-packet inference is effective in improving the RFFI performance in low SNR conditions.

\section*{Acknowledgement}
The work was in part supported by the UK Royal Society Research Grants under grant ID RGS\slash R1\slash 191241 and the national key research and development program of China under grant ID 2020YFE0200600.

\bibliographystyle{IEEEtran}
\bibliography{IEEEabrv,mybibfile}

\end{document}